\documentclass{SSBAtran}

\usepackage{graphicx}
\usepackage[cmex10]{amsmath}

\begin{document}
\title{Combining Vision, Machine Learning and Automatic Control  to Play the \\ Labyrinth Game}

% If one affiliation:
\author{\IEEEauthorblockN{%
Kristoffer \"Ofj\"all and
Michael Felsberg}
\IEEEauthorblockA{Computer Vision Laboratory, Department of E.E. Link\"oping University, Sweden\\
Email: $\{$kristoffer.ofjall, michael.felsberg$\}$@liu.se}
}

\maketitle

\begin{abstract}
The labyrinth game is a simple yet challenging platform, not only for humans but also for control algorithms and systems. The game is easy to understand but still very hard to master. From a system point of view, the ball behaviour is in general easy to model but close to the obstacles there are severe non-linearities. Additionally, the far from flat surface on which the ball rolls provides for changing dynamics depending on the ball position.

The general dynamics of the system can easliy be handled by traditional automatic control methods. Taking the obstacles and uneaven surface into accout would require very detailed models of the system. A simple deterministic control algorithm is combined with a learning
control method.  The simple control method provides initial training data. As the learning method is trained, the system can learn from the results of its own actions and the performance improves well beyond the performance of the initial controller.

A vision system and image analysis is used to estimate the ball position while a combination of a PID controller and a learning controller based on LWPR is used to learn to navigate the ball through the maze.

%simple yet challenging platform for evaluation of realtime control algorithms. game harder than it may seem, stopping at obstacles (non-contionus), uneaven maze (dents, dust (presentation: how many have blamed poor performance on dusty maze?). Position dependent behavior.

%connection between feel of flight and mastering the labyrinth game

%Some other less abstract abstract.

%The evaluated learning systems used traditional control algorithms to provide initial training data. After initial training, the systems learned from their own actions and after a while they outperformed the controller used to provide initial training.

\end{abstract}

% Don't add keywords to the paper!

\section{Introduction}
The \textsc{brio} labyrinth has challenged humans since~\oldstylenums{1946}. The objective is simple: guide the ball through the maze by tilting the plane while avoiding the holes. Most people who have tried it can tell that in practice, the game is really not that simple. By means of computer vision and servo actuators, the challenge can now be handed over to the machines with the same premises as human players. 

A platform for evaluation of control algorithms has been created. The controlling system has to determine the correct action solely based on the visual appearance of the game and the knowledge of previous control signals. Building an evaluation system based on the labyrinth game enables humans to easily relate to the performance of the evaluated control strategies.

\begin{figure}[t]
\centering
\includegraphics[width=1.0\columnwidth]{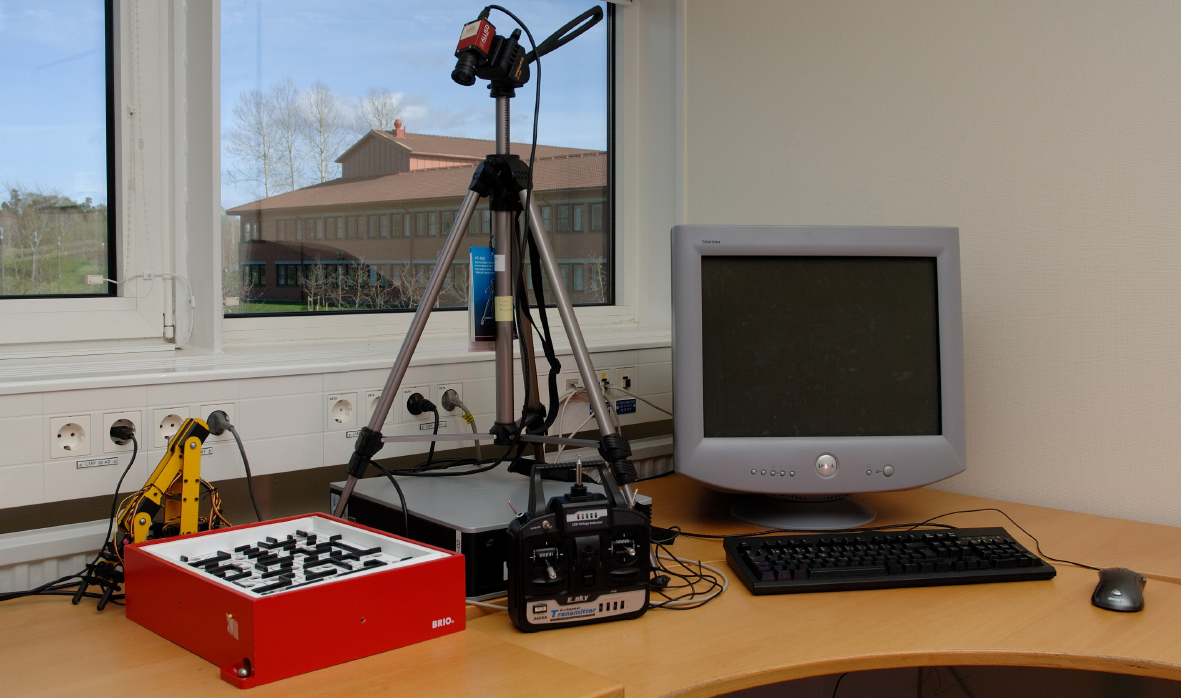}\\
\caption{The system.}
\label{fig_systemOverview}
\end{figure}

An overview of the physical system is provided in Fig.~\ref{fig_systemOverview}. A short description of the implemented and evaluated control strategies is provided in section~\ref{sec_sysSetup}. The evaluation is presented in section~\ref{sec_eval} and conclusions in section~\ref{sec_conclusions}. A more detailed description of the system is available in \cite{Ofjall10}.

\section{System Setup}
\label{sec_sysSetup}

\subsection{Controllers}
For evaluation purposes, three different control strategies have been implemented. These are designated \textsc{pid}, \textsc{lwpr}-2 and \textsc{lwpr}-4.

All strategies uses the same deterministic path planning, a desired ball position is selected from a fixed path depending on the current ball position.

\subsubsection{\textsc{Pid}}
The
proportional-integral-derivative controller 
\begin{equation}
u(t) = 
Pe(t) +
D\frac{de(t)}{dt} +
I\int_0^t e(\tau)\ \mathrm{d}\tau
\end{equation}
is the foundation of classical control theory where $u(t)$ is the control signal and $e(t)$ is the control error.
The parameters $P$, $I$ and $D$ are used to adjust the influence of the proportional part, the derivative part and the integrating part respectively.
Hand tuned dual \textsc{pid} controllers, one for each maze dimension, is used in the system.

\subsubsection{Learning Controllers}
The learning controllers, \textsc{lwpr}-2 and \textsc{lwpr}-4, uses Locally Weighted Projection Regression, \textsc{lwpr} 
\cite{Vijayakumar00locallyweighted}, to learn the inverse dynamics of the system. \textsc{Lwpr} uses several local linear models weighted together to form the output. The parameters of each local model is adjusted online by a modified partial least squares algorithm. The size and number of local models are also adjusted online depending on the local structure of the function to be learned.

For a time discrete system, the inverse dynamics learning problem can be stated as learning the mapping
\begin{equation}\label{learning:statemapping}
\left(\begin{array}{l}
\mathbf{x} \hfill \\
\mathbf{x}^+ \hfill
\end{array}\right)
\rightarrow
\begin{pmatrix}
\mathbf{u}
\end{pmatrix} \qquad.
\end{equation}
Consider a system currently in state $\mathbf{x}$, applying a control signal $\mathbf{u}$ will put the system in another state $\mathbf{x}^+$. Learning the inverse dynamics means that given the current state $\mathbf{x}$ and a desired state $\mathbf{x}^+$, the learning system should be able to estimate the required control signal $\mathbf{u}$ bringing the system from $\mathbf{x}$ to $\mathbf{x}^+$. 

%In the evaluation, two different choices of the state vector $\mathbf{x}$ are compared. One is based on velocities and the other uses both velocities and the absolute position of the ball. The first option is similar to the \textsc{pid} with respect to what kind of information is available to the controlling algorithm.

The desired state of the game is expressed as a desired velocity of the ball in all the conducted experiments involving learning systems. This desired velocity has a constant speed and is directed towards the point selected by the path planner.
The learning systems are trained online. The current state and a desired state is fed into the learning system and the control signal is calculated. When the resulting state of this action is known, the triple previous state, applied control signal and the resulting state is used for training. The learning systems are thus able to learn from their own actions. 

In the cases where the learning system is unable to make control signal predictions due to lack of training data in the current region, the \textsc{pid} controller is used instead. The state and control signal sequences generated by the \textsc{pid} is used as training data for the learning system. Thus, when starting an untrained system, the \textsc{pid} controller will control the game completely. As the learning system gets trained, control of the game will be handled by the learning system to a greater and greater extent.

In the following expressions, $p$, $v$ and $u$ denote position, velocity and control signal respectively. In Eqs.~\eqref{learning:splitNoPosition1} and~\eqref{learning:splitPosition1} a subscript $o$ or $i$ indicates if the aforementioned value correspond to the direction of tilt for the outer or inner gimbal ring of the game.
%TODO: explain inner/outer gimbal

\subsubsection{\textsc{Lwpr}-2}
The \textsc{lwpr}-2 controller tries to learn the mappings
\begin{equation}\label{learning:splitNoPosition1}
\begin{matrix}
\begin{pmatrix}
v_o \hfill \\
v_o^+  \hfill \\
\end{pmatrix}
\rightarrow
\begin{pmatrix}
u_o
\end{pmatrix} ,
\qquad
\begin{pmatrix}
v_i \hfill \\
v_i^+  \hfill \\
\end{pmatrix}
\rightarrow
\begin{pmatrix}
u_i
\end{pmatrix} \qquad.
\end{matrix}
\end{equation}
This setup makes the same assumptions regarding the system as those made for the \textsc{pid} controller. First, the ball can not behave differently in different parts of the maze. Secondly, the outer servo should not affect the ball position in the inner direction and vice versa.

\subsubsection{\textsc{Lwpr}-4}
By adding the absolute position to the input vectors, \textsc{lwpr}-\oldstylenums{4} is obtained. The mappings are
\begin{equation}\label{learning:splitPosition1}
%\begin{matrix}
\begin{pmatrix}
v_o \hfill \\
p_o \hfill \\
p_i \hfill \\
v_o^+  \hfill \\
\end{pmatrix}
\rightarrow
\begin{pmatrix}
u_o
\end{pmatrix} ,
\qquad
\begin{pmatrix}
v_i \hfill \\
p_o \hfill \\
p_i \hfill \\
v_i^+  \hfill \\
\end{pmatrix}
\rightarrow
\begin{pmatrix}
u_i
\end{pmatrix}
%\end{matrix}  
\qquad .
\end{equation}

This learning system should have the possibility to handle different dynamics in different parts of the maze. Still it is assumed that the control signal in one direction has little effect on the ball movement in the other.

\subsection{Vision and Image Processing}
Vision is the only means for feedback available to the controlling system. The controller is dependent on knowing the state of the ball in the maze. The ball position, in a coordinate system fixed in the maze, is estimated by means of a camera system and a chain of image processing steps.

\begin{figure}[t]
\centering
\includegraphics[width=1.0\columnwidth]{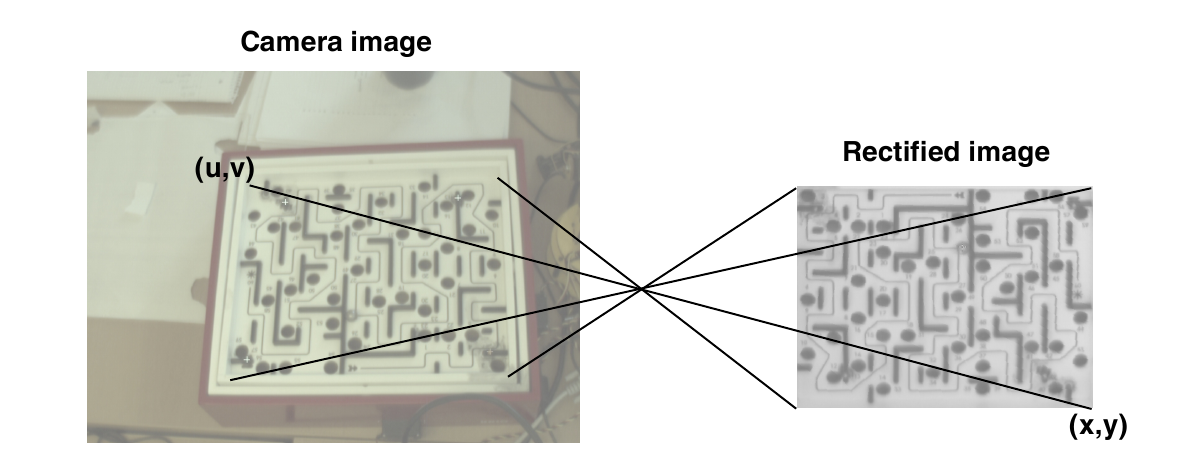}\\
\caption{Rectifying homography.}
\label{fig_homographyRect}
\end{figure}

The maze is assumed to be planar and the lens distortion is negliable
 so the mapping between image coordinates and maze coordinates can be described by a homography, Fig.~\ref{fig_homographyRect}. To simplify homography estimation, four colored markers with known positions within the maze are detected and tracked.

\begin{figure}[t]
\centering
\includegraphics[width=1.0\columnwidth]{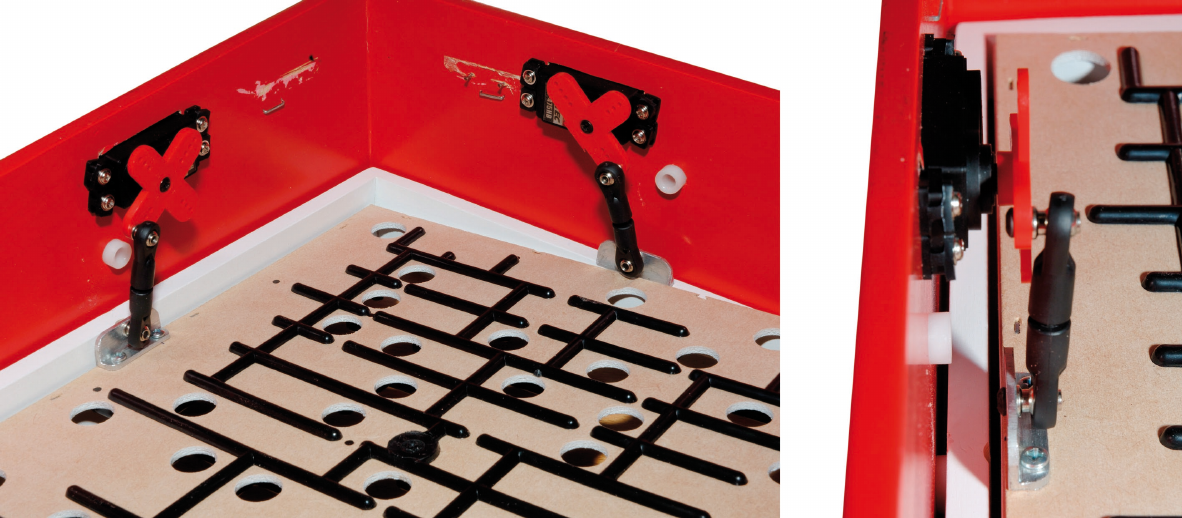}\\
\caption{Servo installation.}
\label{fig_servoInstallation}
\end{figure}

As the maze is stationary in the rectified images even when the maze or camera is moved, a simple background model and background subtraction can be used to find the position of the ball. An approximate median background model, described in \cite{ardoPHDbgModels}, is used. After background subtraction and removal of large differences originating from the high contrast between the white maze and the black obstacles, the ball position is easily found.

\subsection{State Estimation}
The ball velocity is needed by the controllers.
%Knowing the ball velocity could be expected to greatly improve the accuracy of the learning controller and is necessary for the \textsc{pid}-controller. 
Direct approximation of the velocity with difference methods provides estimations drowned in noise. A Kalman filter \cite{kalman} is used to filter the position information as well as to provide an estimate of the ball velocity.

A time discrete Kalman filter is used, based on a linear system model
\begin{equation}
\begin{array}{llllllll}
\mathbf{x}_{n+1} & = & \mathbf{A}\mathbf{x}_n & + & \mathbf{B}\mathbf{u}_{n} & + & \mathbf{w}_{n} \\
\mathbf{y}_{n} & = & \mathbf{C}\mathbf{x}_{n} & + & \mathbf{v}_n & & & ,
\end{array}
\end{equation}
with state vector $\mathbf{x}_n$ at time $n$, output $\mathbf{y}_n$, control signal $\mathbf{u}_n$, process noise $\mathbf{w}_n$, measurement noise $\mathbf{v}$ and system parameters $\mathbf{A}$, $\mathbf{B}$, $\mathbf{C}$.

\subsubsection{Linear System Model}
The servo is modeled as a proportionally controlled motor with a gearbox. The servo motor (\textsc{dc}-motor) and gearbox is modeled as 
\begin{equation} \label{hwmodel:dcmotor}
\ddot{\theta} = -a\dot{\theta} + bv
\end{equation}
 where $\theta$ is the output axis angle and $v$ is the input voltage. The internal proportional feedback $v=K(K_2u-\theta)$, where $u$ is the angular reference signal, yields the general second order system
\begin{equation} \label{hwmodel:dcservo}
\ddot{\theta} = -bK\theta -a\dot{\theta} + bKK_2u \qquad.
\end{equation}

The physical layout of the control linkage provides for an approximate offset linear relation between servo deflection, maze tilt angle and ball acceleration.
Thus, the ball motion could be modeled as
\begin{equation} \label{hwmodel:ballmotion}
\ddot{y} = c(\theta + \theta_0) - d\dot{y} 
\end{equation}
as long as the ball avoids any contact with the obstacles.
% where $y$ is the ball position and $\theta_0$ is the maze offset. In the equation, $d$ model friction and $c$ is the linearized relation between maze tilt and ball acceleration. Both parameters contain the ball mass.

Using the state vector
$\mathbf{x} = \begin{pmatrix} y & \dot{y} & \theta & \dot{\theta} & \theta_0 \end{pmatrix} ^\mathrm{T} $
%\begin{equation} \label{hwmodel:state}
%\mathbf{x} =
%\begin{pmatrix} x_1 \\ x_2 \\ x_3 \\ x_4 \\ x_5 \end{pmatrix} =
%\begin{pmatrix} y \\ \dot{y} \\ \theta \\ \dot{\theta} \\ \theta_0 \end{pmatrix}
%\end{equation}
 the combination of equations~\eqref{hwmodel:dcservo} and~\eqref{hwmodel:ballmotion} can be expressed as the continuous time state space model
\begin{equation} \label{hwmodel:modelcont}
\begin{matrix}
\dot{\mathbf{x}} & = & 
\begin{pmatrix}
0 & 1 & 0 & 0 & 0 \\
0 & -d & c & 0 & c \\	
0 & 0 & 0 & 1 & 0 \\	
0 & 0 & -bK & -a & 0 \\	
0 & 0 & 0 & 0 & 0 \\	
\end{pmatrix} \mathbf{x}
& + &
\begin{pmatrix}
0 \\ 0 \\ 0 \\ bKK_2 \\ 0
\end{pmatrix} u
\\
\\
y & = & \begin{pmatrix} 1&0&0&0&0 \end{pmatrix} \mathbf{x} &  .
\end{matrix}
\end{equation}

A time discrete model can be obtained using forward difference approximations of the derivatives $\dot{x} \approx \frac{x^{n+1}-x^n}{T}$ $\Leftrightarrow$ $x^{n+1} \approx x^n + T\dot{x}$ where $T$ is the sampling interval.
Using standard methods for system identification, \cite{mosboken}, the unknown parameters can be identified.

\subsection{Actuators}
For controlling the maze, two standard servos for radio controlled models have been installed in the game, see Fig.~\ref{fig_servoInstallation}.

\begin{figure}[t]
\centering
\includegraphics[width=0.8\columnwidth]{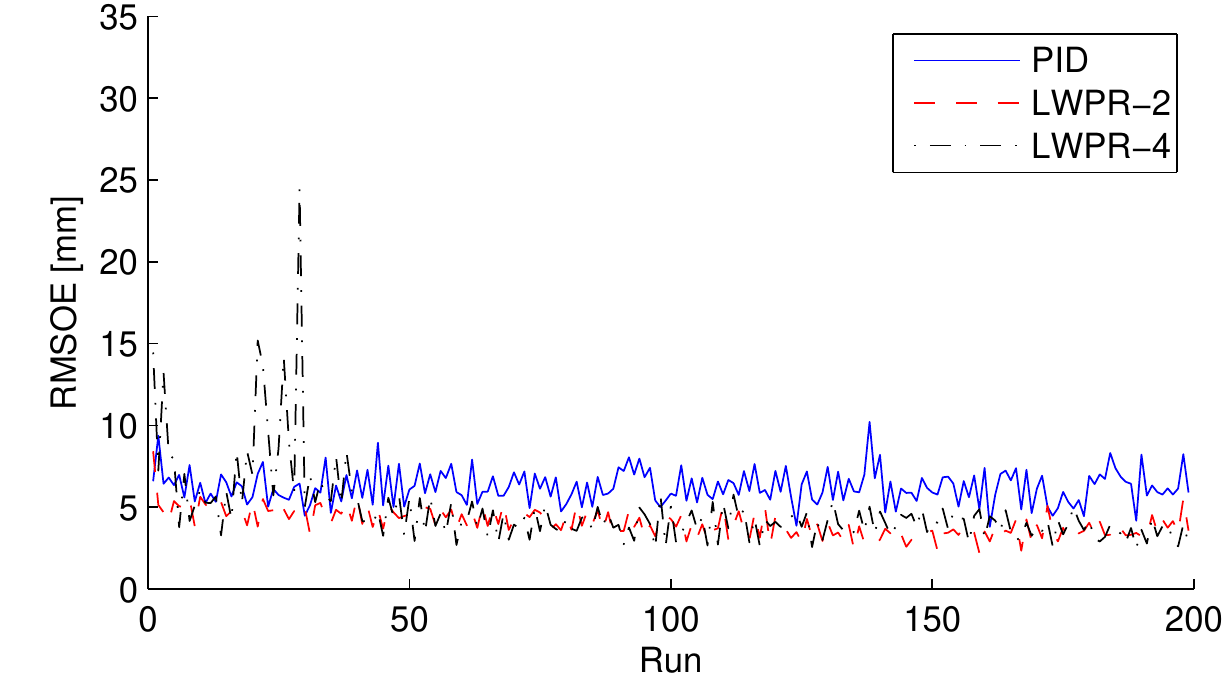}\\
\caption{Deviation from desired path, scenario 1.}
\label{fig_resScen1}
\end{figure}

\begin{figure}[t]
\centering
\includegraphics[width=0.8\columnwidth]{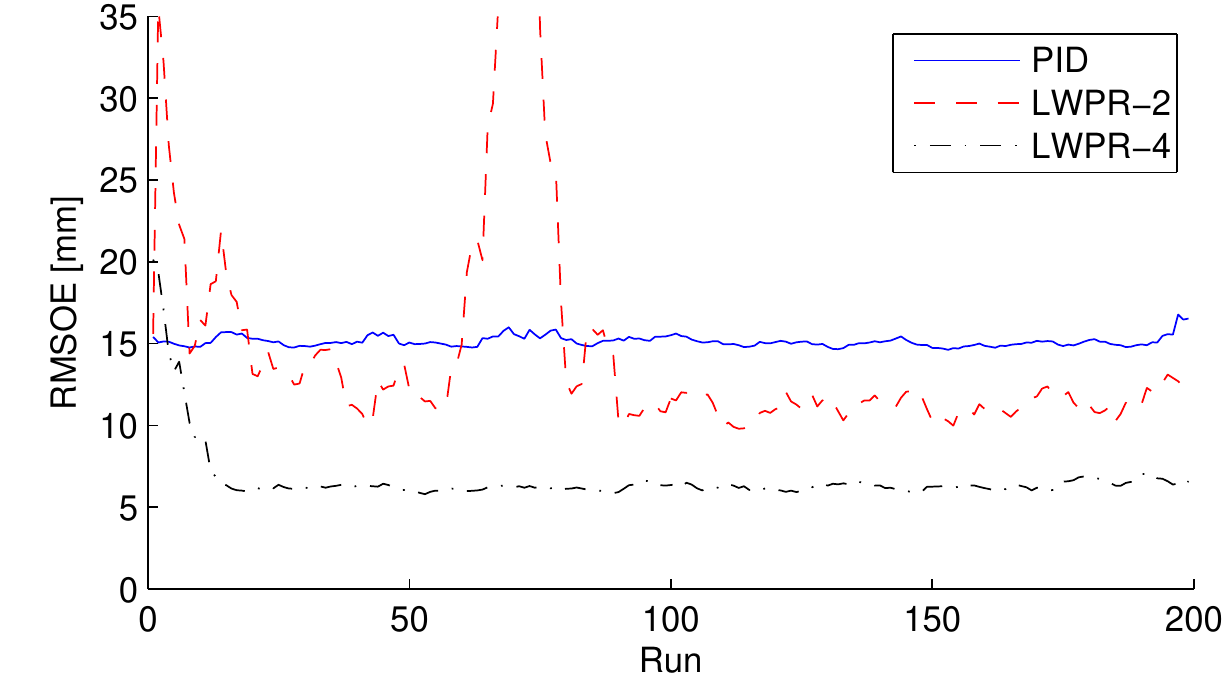}\\
\caption{Deviation from desired path, smoothed over runs, scenario 2a.}
\label{fig_resScen2a}
\end{figure}

\section{Evaluation}
\label{sec_eval}
To facilitate more fine grained performance measurements, a different maze is used for evaluation. The alternative maze is flat and completely free of holes and obstacles. The controllers are evaluated by measuring the deviation from a specified path. \textsc{Rmsoe} is the root mean squared orthogonal deviation of the measured ball positions from the desired path. The \textsc{rmsoe} averaged over runs 171 to 200 for each scenario and controller is shown in Table~\ref{tab_totalResults}.

\subsection{Scenario 1}
The first scenario is a simple sine shaped path. The deviation from the desired path for the three different controllers are shown in Fig.~\ref{fig_resScen1}. The learning controllers are started completely untrained and after some runs they outperform the \textsc{pid} controller used to generate training data initially. As expected, the pure \textsc{pid} controller has a constant performance over the runs.

\begin{figure}[t]
\centering
\includegraphics[width=0.8\columnwidth]{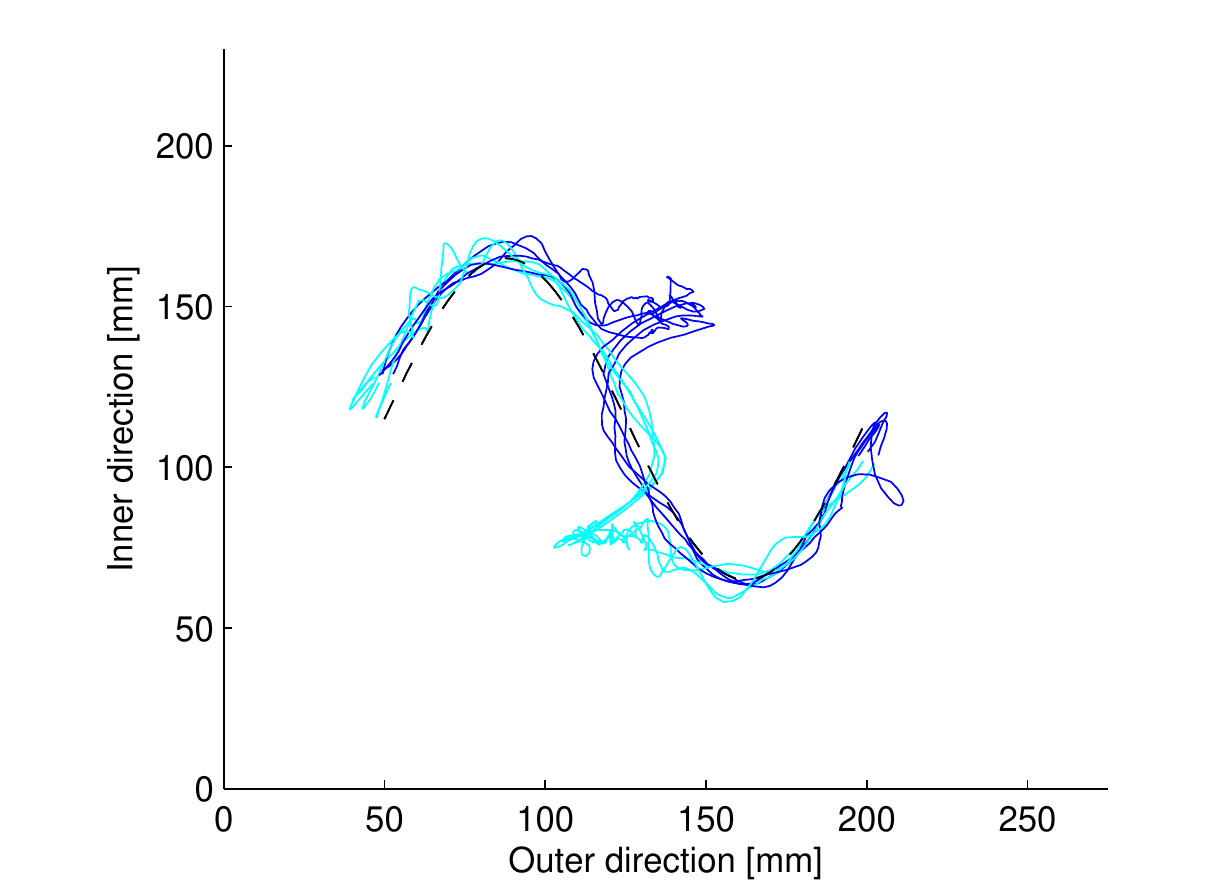}\\
\caption{Eight runs by the \textsc{pid} controller in scenario 2a. Cyan lines indicate forward runs, blue lines are used for reverse runs. The dashed black line is the desired trajectory.}
\label{fig_scen2aPidrun}
\end{figure}

\begin{figure}[t]
\centering
\includegraphics[width=1.0\columnwidth]{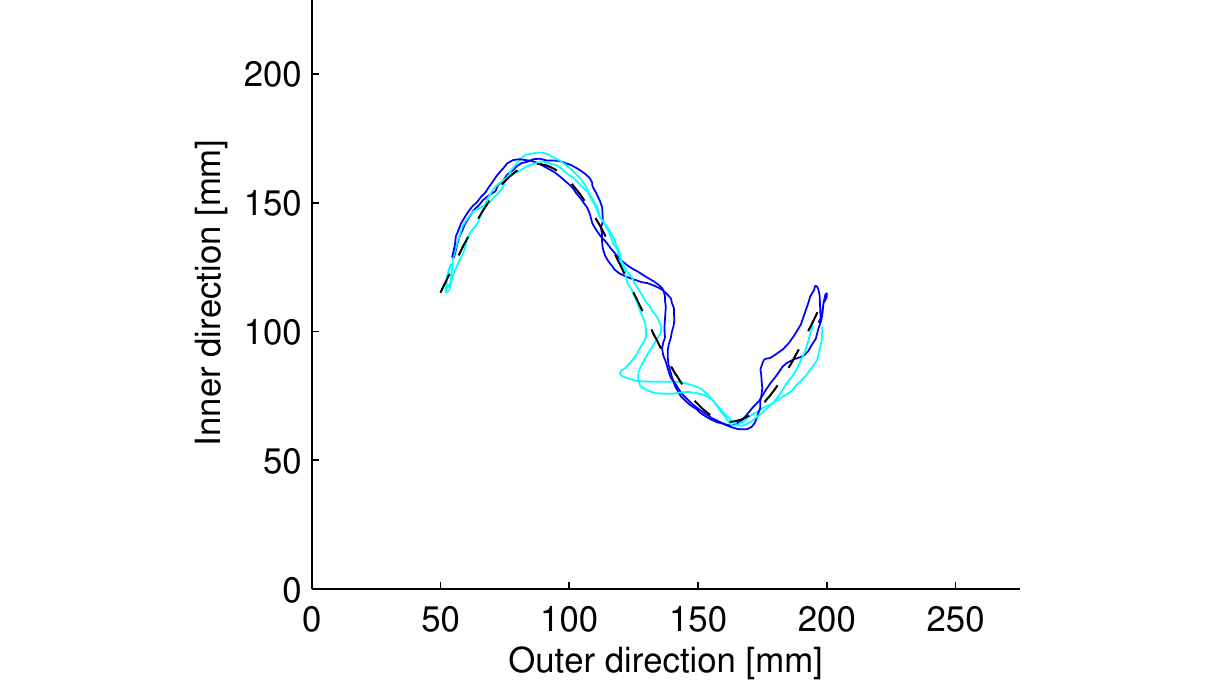}\\
\caption{Four runs (200 to 203) by the \textsc{lwpr}-4 in scenario 2b. Cyan lines indicate forward runs, blue lines are used for reverse runs. The dashed black line is the desired trajectory.}
\label{fig_scen2bLwprRun}
\end{figure}

\subsection{Scenario 2}
The desired path for the second scenario is the same as for the first. In the second scenario, the game dynamics are changed depending on the position of the ball. In scenario 2a, a constant offset is added to the outer gimbal servo signal when the ball is in the bottom half of the maze. In scenario 2b, the outer gimbal servo is reversed when the ball is in the bottom half of the maze.

The deviation for scenario 2a is shown in Fig.~\ref{fig_resScen2a}. As expected, the position dependent \textsc{lwpr}-4 controller performs best. A few runs by the \textsc{pid} controller in scenario 2a is shown in Fig.~\ref{fig_scen2aPidrun}. The effect of the position dependet offset is clear. The integral term need some time to adjust after each change of half planes.

Only the \textsc{lwpr}-4 controller is able to control the ball in scenario 2b, the two other controllers both compensate in the wrong direction. In this scenario, the \textsc{pid} controller can not be used to generate training data. For this experiment, initial training data was generated by controlling the game manually. The position dependent control reversal was hard to learn even for the human subject. A few runs by \textsc{lwpr}-4 is shown in Fig.~\ref{fig_scen2bLwprRun}.

\subsection{Scenario 3}
The desired path for scenario 3 is the path of the real maze. In this scenario, only \textsc{lwpr}-4 was able to handle the severe nonlinearities close to the edges of the maze. The other two controllers were prone to oscillations with increasing amplitude.
Still, the \textsc{pid} controller was useful for generating initial training data as the initial oscillations were dampened when enough training data had been collected. These edge related problems illustrates why only \textsc{lwpr}-4 was able to control the ball in the real maze with obstacles.

\begin{figure}[t]
\centering
\includegraphics[width=0.8\columnwidth]{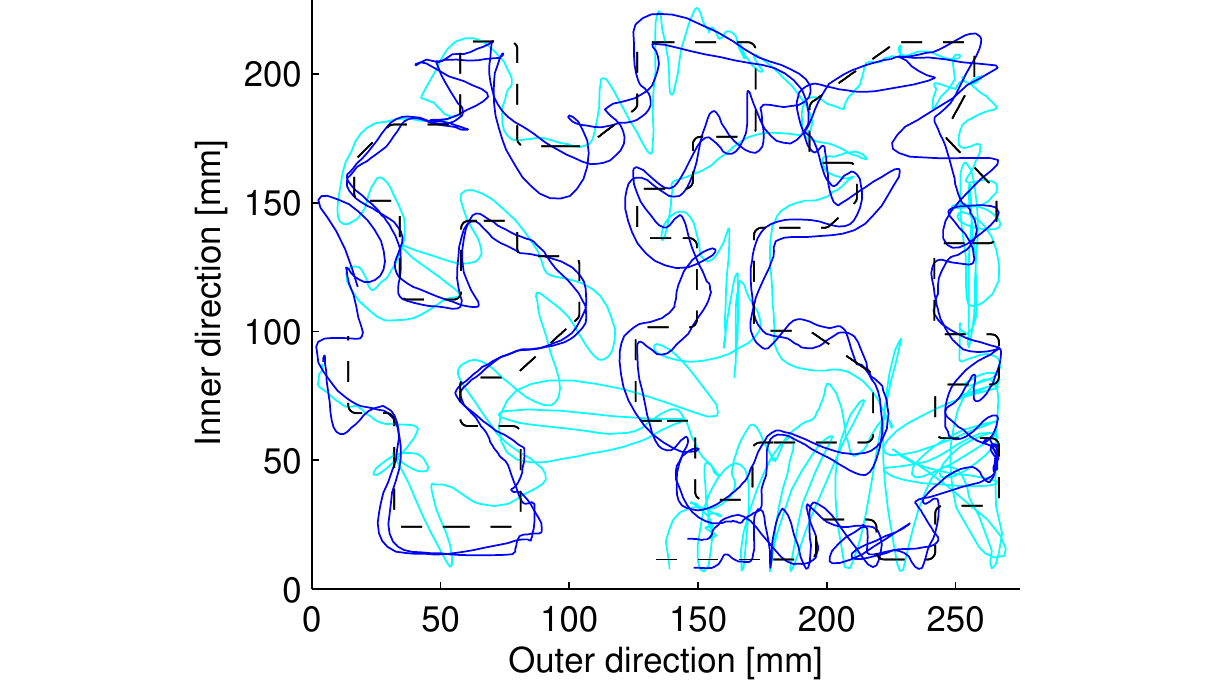}\\
\caption{Trajectories from early runs by \textsc{lwpr}-4 in scenario 3. Cyan lines indicate forward runs, blue lines are used for reverse runs. The dashed black line is the desired trajectory.}
\label{fig_scen3early}
\end{figure}

\begin{figure}[t]
\centering
\includegraphics[width=0.8\columnwidth]{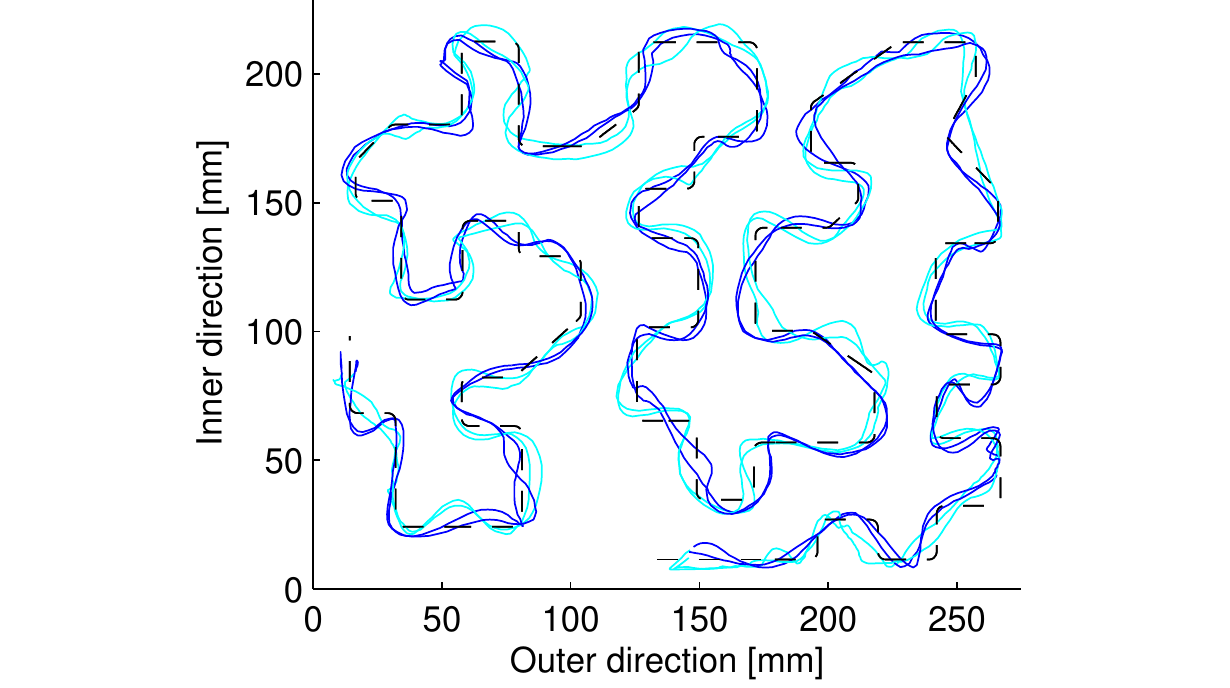}\\
\caption{Trajectories from late runs by \textsc{lwpr}-4 in scenario 3. Cyan lines indicate forward runs, blue lines are used for reverse runs. The dashed black line is the desired trajectory.}
\label{fig_scen3late}
\end{figure}

Some early runs are shown in Fig.~\ref{fig_scen3early}, the oscillations from the \textsc{pid} controller can clearly be seen. Some later runs are shown in Fig.~\ref{fig_scen3late}. The remaining tendency to cut corners can to some extent be explained by the path planning algorithm.

\section{Conclusions}
\label{sec_conclusions}
Both \textsc{lwpr} based controlling algorithms outperform the \textsc{pid} in all scenarios. From this, two conclusions may be drawn. First, it should be possible to design a much better traditional controller. Secondly, by learning from their own actions, the learning systems are able to perform better than the controlling algorithm used to provide initial training data.

The \textsc{lwpr}-4 requires more training data than \textsc{lwpr}-2. According to the authors of \cite{Vijayakumar00locallyweighted}, this should not necessarily be the case. However, depending on the initial size of the local models, more local models are needed to fill a higher dimensional input space.
%TODO: ev higher dimensional input space

\newpage

Finally, the combination of a simple deterministic controller and a learning controller has been powerful. Designing a better deterministic controller would require more knowledge of the system to be controlled, which may not be available. A learning controller requires training data before it is useful. Combining a learning controller with a simple deterministic controller, the control performance start at the level of the simple controller and is improved as the system is run by automatic generation of training data.

%\subsection{Table of numbers}
%Table of numerical results

\begin{table}[t]
\caption{Mean \textsc{rmsoe} for 30 runs in the end of each scenario. The standard deviations are given within parentheses.}
\label{tab_totalResults}
    \begin{tabular}{ l | r@{\ }l  r@{\ }l  r@{\ }l }
      & \multicolumn{2}{c}{\textsc{pid}} & 
        \multicolumn{2}{c}{\textsc{lwpr}-\oldstylenums{2}} & 
        \multicolumn{2}{c}{\textsc{lwpr}-\oldstylenums{4}} \\ \hline
  Scenario 1  & 6.0  & (0.8) & 3.5  & (0.5) & 3.7 & (0.9)   \\
  Scenario 2a & 15.5 & (1.6) & 11.5 & (1.8) & 6.5 & (1.6)   \\
  Scenario 2b & \multicolumn{2}{c}{DNF} & \multicolumn{2}{c}{DNF} & 5.6 & (1.1)   \\
  Scenario 3  & \multicolumn{2}{c}{DNF} & \multicolumn{2}{c}{DNF} & 3.8 & (0.6)   \\
  \end{tabular}
\end{table}

%\newpage % this command is to balance the two columns somewhat. If the
         % column break has to come half-way the reference list, use
         % \IEEEtriggeratref{} command explained below.

\section*{Acknowledgment}
The authors would like to thank Fredrik Larsson for inspiration and discussions.
This research has received funding from the EC's 7th Framework Programme (FP7/2007-2013), grant agreement 247947 (GARNICS).

%\vspace{2mm} % this command adds a little vertical space to fine-tune
             % the column balancing. LaTeX has no mechanism to balance
             % columns.



\bibliographystyle{SSBAtrans}
\bibliography{bibl}

\end{document}